\documentclass[11pt]{article}
\usepackage[margin=1in]{geometry}
\usepackage{amsmath,amssymb,amsthm,mathtools}
\usepackage[shortlabels]{enumitem}
\usepackage{microtype}
\usepackage{xcolor}
\usepackage[colorlinks=true,linkcolor=blue!50!black,citecolor=green!40!black,urlcolor=blue!50!black]{hyperref}

\numberwithin{equation}{section}
\newtheorem{theorem}{Theorem}[section]
\newtheorem{proposition}[theorem]{Proposition}
\newtheorem{lemma}[theorem]{Lemma}
\newtheorem{corollary}[theorem]{Corollary}
\theoremstyle{definition}
\newtheorem{definition}[theorem]{Definition}
\newtheorem{example}[theorem]{Example}
\theoremstyle{remark}
\newtheorem{remark}[theorem]{Remark}

\newcommand{\Hil}{\mathcal{H}}
\newcommand{\Ps}{\mathcal{P}}
\newcommand{\Pphy}{\mathcal{P}_{\rm phy}}
\newcommand{\Sphy}{\mathcal{S}_{\rm phy}}
\newcommand{\Cq}{\mathcal{C}}
\newcommand{\Mop}{\hat{\mathbf{M}}}
\newcommand{\Mcl}{\mathbf{M}}
\newcommand{\Gb}{\bar{\Gamma}}
\newcommand{\Gh}{\hat{\Gamma}}
\newcommand{\Lie}{\mathfrak{g}}
\newcommand{\ket}[1]{|#1\rangle}

\newcommand{\ip}[2]{\langle #1|#2\rangle}
\newcommand{\ev}[1]{\langle #1\rangle}
\newcommand{\Pz}{\hat{P}_0}
\newcommand{\CP}{\mathbb{C}P}
\newcommand{\R}{\mathbb{R}}
\newcommand{\C}{\mathbb{C}}
\newcommand{\Z}{\mathbb{Z}}
\newcommand{\eps}{\varepsilon}
\DeclareMathOperator{\grad}{grad}
\DeclareMathOperator{\Hess}{Hess}
\DeclareMathOperator{\supp}{supp}
\DeclareMathOperator{\spec}{spec}
\DeclareMathOperator{\Crit}{Crit}

\title{On the geometry of quantum constrained systems\\[4pt]
}
\author{Alejandro Corichi\\[2pt]
\small Centro de Ciencias Matem\'aticas, Universidad Nacional Aut\'onoma de M\'exico,\\
\small UNAM-Campus Morelia, Morelia, Michoac\'an, Mexico}
\date{}

\begin{document}
\maketitle

\begin{abstract}
Geometric methods have proved useful in the Hamiltonian description of classical constrained systems. We develop the corresponding geometry for quantum constrained systems, using the formulation of quantum mechanics in which the space of rays $\Ps$ is a K\"ahler manifold and unitary transformations are its canonical transformations. The Dirac conditions $\hat C_i\Psi=0$ define the zero set of a single function $\Cq=\ev{\Mop}$, the expectation value of the master constraint operator. Our central identity is
$\Cq=|\mu|^2+\tfrac12\|X_\mu\|^2$, where $\mu=(\ev{\hat C_i})$ is the moment map of the gauge action on $\Ps$ and $X_\mu$ its fundamental vector fields: the quantum condition imposes the classical-type condition $\mu=0$ \emph{and} the collapse of the gauge orbits to points. This explains, rather than merely observes, why no second reduction is needed. The physical state space is a totally geodesic K\"ahler submanifold lying in the deepest stratum of the K\"ahler quotient $\Ps/\!\!/G$; $\Cq$ is a Morse--Bott function whose transverse Hessian is the non-zero spectrum of $\Mop$. Its gradient flow is heat-kernel-smeared group averaging, converges to the nearest-point projection onto physical states, and its fibres are invariant under the complexified gauge group. Physical observables are characterised to all orders by the double-commutator condition, which is equivalent to $[\hat O,\hat P_0]=0$, with $\hat P_0$ the projector onto physical states. For constraints with a central term, including second-class pairs, we derive the bound $\Cq_1+\Cq_2\ge|\{c_1,c_2\}|$, whose minimisers implement the Gupta--Bleuler condition. When zero lies in the continuous spectrum, the gradient flow converges to Dirac states in the sense of Grundling and Hurst, and its large-time asymptotics reproduce the rigging map of refined algebraic quantization. Finally, pulled back along coherent states, $\Cq$ equals the classical master constraint plus an $O(\hbar)$ term given by the squared speed of the classical gauge orbits. Together these results connect the Dirac procedure, the master constraint programme, refined algebraic quantization and symplectic reduction within one geometric picture.
\end{abstract}


\tableofcontents

\section{Introduction}

Most fundamental theories of physics, when analysed from the Hamiltonian perspective, are subject to constraints $C_i\approx0$ among their canonical variables. Notable examples are totally constrained systems, whose Hamiltonian is a combination of constraints and whose time evolution is therefore also generated by constraints, such as general relativity on a compact Cauchy surface. The description of classical constrained systems has benefited greatly from symplectic geometry. Dirac's classification into first- and second-class constraints is a statement about the degeneracy of the symplectic form restricted to the constraint surface; gauge-invariant information is encoded in the quotient by the orbits of the first-class constraints \cite{HT,Kuchar}. It is natural to ask whether this geometric understanding extends to the quantum theory.

There are two main strategies for quantizing constrained systems: Dirac quantization and reduced phase space quantization. Within the Dirac approach several implementations have been developed, among them refined algebraic quantization (RAQ) \cite{Higuchi,ALMMT,Marolf}, coherent-state and projection-operator methods \cite{Klauder}, and the master constraint programme \cite{Thiemann}. In the Dirac approach the constraints are promoted to operators $\hat C_i$ on a kinematical Hilbert space and physical states are those satisfying
\begin{equation}
\hat C_i\,\Psi=0. \label{eq:dirac}
\end{equation}
The first difference between the classical and quantum treatments is obvious. Classically, reduction takes two steps: restriction to the constraint surface, followed by the quotient along gauge orbits. Quantum mechanically there is a single condition. It is often said that this one quantum step is equivalent to the two classical ones. The purpose of this paper is to make that statement precise, and to see what else the geometry teaches us.

We use the geometric formulation of quantum mechanics \cite{GQM}, in which the space of rays $\Ps$ is a K\"ahler manifold. Its symplectic structure implements unitary transformations as canonical transformations, while its Riemannian metric encodes uncertainties and transition probabilities. In an earlier version of this work \cite{Corichi2008} we observed that the Dirac conditions can be written, as in the classical case, as the zero set of a function on $\Ps$, and that the condition for an operator to be a physical observable leads naturally to the double commutator of the master constraint programme. The present version replaces several of those observations by theorems and extends the analysis to anomalous, second-class and continuous-spectrum constraints. The main results are the following.

\begin{enumerate}[(i)]
\item \emph{Decomposition theorem} (Theorem~\ref{thm:decomp}). The quantum constraint function $\Cq=\ev{\Mop}$ splits as $\Cq=|\mu|^2+\tfrac12\|X_\mu\|^2$: a moment-map term mirroring the classical constraints and the squared length of the gauge vector fields. Hence $\Pphy=\mu^{-1}(0)\cap\Ps^G$: the quantum condition does not quotient the gauge orbits, it retains only the points whose orbit is a single point.
\item \emph{Geometry of $\Pphy$} (Propositions~\ref{prop:Pphy}, \ref{prop:morsebott}, \ref{prop:kahlerred}). $\Pphy$ is a totally geodesic K\"ahler submanifold with no degenerate directions, and it is the fixed-point stratum of the K\"ahler quotient $\Ps/\!\!/G$. The function $\Cq$ is Morse--Bott, an inadmissible description of $\Pphy$ exactly like the classical master constraint, and its transverse Hessian is the spectrum of $\Mop$.
\item \emph{The quantum projection} (Theorem~\ref{thm:flow}). The gradient flow of $\Cq$ is normalized imaginary-time evolution under $\Mop$, equals heat-kernel-smeared group averaging, and converges to the nearest-point projection $\eta$ onto $\Pphy$. The fibres of $\eta$ are complex affine spaces invariant under the complexified gauge group; this is the precise sense in which the quantum quotient differs from the classical one.
\item \emph{Observables} (Proposition~\ref{prop:obs}). The double-commutator condition is not merely the first term of an expansion; on physical states it is equivalent to $[\hat O,\Pz]=0$, to all orders. Every kinematical operator induces a physical observable by compression.
\item \emph{Anomalies and second-class constraints} (Theorem~\ref{thm:anomaly}). The geometric uncertainty bound $\Cq_1+\Cq_2\ge c_1^2+c_2^2+|\{c_1,c_2\}|$ shows that a central term is a symplectic obstruction to Dirac states. The minimisers of $\Cq$ for a second-class pair are exactly the Gupta--Bleuler states.
\item \emph{Continuous spectrum} (Theorems~\ref{thm:dirac-states}, \ref{thm:rigging}). The gradient flow converges weak-$*$ to Dirac states, and $\eta_{\rm RAQ}(\psi)[\phi]=\lim_{t\to\infty}(4\pi t)^{k/2}\ip{\phi}{e^{-t\Mop}\psi}$.
\item \emph{Semiclassical limit} (Proposition~\ref{prop:semiclassical}). On coherent states, $\Cq=C^2+\tfrac{\hbar}{2}\bigl(|\nabla C|^2+C\,\Delta C\bigr)+O(\hbar^2)$: the classical master constraint plus half the squared gauge-orbit speed.
\end{enumerate}

The structure of the paper is as follows. Section~\ref{sec:classical} reviews classical constrained systems in symplectic language. Section~\ref{sec:gqm} fixes conventions for geometric quantum mechanics and collects the lemmas we need. Section~\ref{sec:first-class} is the core of the paper and treats first-class constraints with zero in the discrete spectrum. Section~\ref{sec:examples} works out two examples. Section~\ref{sec:anomaly} treats anomalies and second-class constraints, Section~\ref{sec:continuous} the continuous spectrum, and Section~\ref{sec:semiclassical} the semiclassical limit. We end with a discussion in Section~\ref{sec:discussion}. Appendix~\ref{app:changes} lists the corrections with respect to the published version \cite{Corichi2008}.

\section{Classical constrained systems}\label{sec:classical}

We recall the treatment of constrained Hamiltonian systems from the perspective of symplectic geometry, assuming the Dirac--Bergmann algorithm \cite{Dirac,HT} has already produced the final set of constraints.

A phase space is a $2N$-dimensional manifold $\Gamma$ with symplectic form $\Omega$. A vector field $V$ generates infinitesimal canonical transformations if $\mathcal L_V\Omega=0$, equivalently, locally, $V^b=\Omega^{ba}\nabla_af=:X_f^b$ for some function $f$; $X_f$ is the Hamiltonian vector field (HVF) of $f$. The Poisson bracket
\begin{equation}
\{f,g\}:=\Omega^{ab}\nabla_af\nabla_bg=\mathcal L_{X_g}f \label{eq:PB}
\end{equation}
gives the change of $f$ along the flow of $g$, and satisfies $X_{\{f,g\}}=-[X_f,X_g]$.

Constraints are $M$ functions $C_i:\Gamma\to\R$; the constraint surface is $\Gb=\{C_i=0\}$. If the gradients $\nabla C_i$ are independent on $\Gb$, the $C_i$ are an \emph{admissible description} of the $(2N-M)$-dimensional surface $\Gb$. We consider first-class systems,
\begin{equation}
\{C_i,C_j\}=F_{ij}{}^kC_k, \label{eq:firstclass}
\end{equation}
so that the HVFs $X_i:=X_{C_i}$ are tangent to $\Gb$, are in involution there, and span the kernel of the pull-back $\bar\Omega$ of $\Omega$ to $\Gb$: for all $\bar V$ tangent to $\Gb$, $\Omega(X_i,\bar V)=\bar V^a\nabla_aC_i=0$. $\Gb$ is thus coisotropic. Taking as gauge transformations those generated by the first-class constraints,\footnote{This is the content of the Dirac conjecture. It is known to fail in certain systems, see e.g.\ \cite{CaboLouis,HT}, and is better regarded as a definition of what we mean by gauge here.} the reduced phase space is the quotient $\Gh=\Gb/\!\sim$ along the orbits of the $X_i$, when this is a manifold. When the $C_i$ have constant structure coefficients $f_{ij}{}^k$ they define a Hamiltonian action of a Lie group $G$ with equivariant moment map $\mu=(C_i):\Gamma\to\Lie^*$, and $\Gh=\mu^{-1}(0)/G$ is the Marsden--Weinstein quotient \cite{MarsdenWeinstein}.

A function $f$ is a (weak) \emph{observable} if the canonical transformations it generates preserve $\Gb$:
\begin{equation}
0=\{C_i,f\}\big|_{\Gb}=-X_i(f)\big|_{\Gb}. \label{eq:classobs}
\end{equation}
Observables are therefore constant along gauge orbits and descend to $\Gh$; since physical distinguishability is measured by observables, points of an orbit are physically indistinguishable.

The \emph{master constraint} \cite{Thiemann} is
\begin{equation}
\Mcl:=K^{ij}C_iC_j, \label{eq:master}
\end{equation}
with $K^{ij}$ constant and positive definite. Clearly $\Mcl^{-1}(0)=\Gb$, but $\Mcl$ is an \emph{inadmissible} description: $\nabla\Mcl$ and $X_{\Mcl}$ vanish on $\Gb$, so $\Mcl$ generates nothing there and $\{\Mcl,f\}|_{\Gb}=0$ for every $f$. To characterise observables one asks instead that the finite flow of $f$ preserve $\Gb$:
\begin{equation}
\alpha^f_t[\Mcl]\big|_{\Gb}:=\sum_{n=0}^\infty\frac{t^n}{n!}\{\Mcl,f\}_{(n)}\Big|_{\Gb}=0,\qquad
\{g,f\}_{(0)}=g,\quad\{g,f\}_{(n+1)}=\{\{g,f\}_{(n)},f\}. \label{eq:finiteflow}
\end{equation}
The first non-trivial condition is the \emph{master equation} $\{\{\Mcl,f\},f\}|_{\Gb}=0$. In fact, since $C_i=0$ on $\Gb$,
\begin{equation}
\{\{\Mcl,f\},f\}\big|_{\Gb}=2K^{ij}\{C_i,f\}\{C_j,f\}\big|_{\Gb}, \label{eq:mastereq}
\end{equation}
which is a positive quadratic form in $\{C_i,f\}$. Thus the master equation alone is equivalent to (\ref{eq:classobs}), and all higher terms of (\ref{eq:finiteflow}) then vanish. We will find an exact quantum analogue of (\ref{eq:mastereq}) in Proposition~\ref{prop:obs}.

\section{Geometric quantum mechanics}\label{sec:gqm}

The geometric formulation of quantum mechanics has been rediscovered several times \cite{GQM}. We fix conventions, correcting a normalization in \cite{Corichi2008}, and collect the facts we need. We set $\hbar=1$ except in Section~\ref{sec:semiclassical}.

Let $\Hil$ be a complex Hilbert space, finite-dimensional for simplicity (the extension to Hilbert manifolds is standard \cite{GQM}), and $\Ps$ its space of rays; for $\dim\Hil=n$, $\Ps\cong\CP^{n-1}$. Viewing $\Hil$ as a real vector space with complex structure $J$ (multiplication by $i$), write
\begin{equation}
\ip{u}{v}=\tfrac12 G(u,v)+\tfrac{i}{2}\Omega(u,v),\qquad \Omega(u,v)=G(Ju,v). \label{eq:hermitian}
\end{equation}
At a unit vector $\psi$ we identify $T_{[\psi]}\Ps$ with the horizontal subspace $\psi^\perp$; $G$, $\Omega$ and $J$ descend to a K\"ahler structure $(g,\Omega,J)$ on $\Ps$ (a multiple of the Fubini--Study structure), with $g(u,v)=2\,{\rm Re}\ip{u}{v}$ and $\Omega(u,v)=2\,{\rm Im}\ip{u}{v}$ for $u,v\in\psi^\perp$.

To each self-adjoint operator $F$ we associate the function $f=\ev F:[\psi]\mapsto\ip{\psi}{F\psi}$ (unit $\psi$), and we write $\tilde F:=F-f([\psi])$. Such functions are precisely those whose HVFs are Killing fields of $g$ \cite{GQM}; we call them \emph{quantum observables} on $\Ps$.

\begin{lemma}[Dictionary]\label{lem:dict}
Let $F,H$ be self-adjoint, $f=\ev F$, $h=\ev H$, and $\psi$ a unit vector. Then:
\begin{enumerate}[(a)]
\item $X_f=-J\tilde F\psi$, and $\grad f=\tilde F\psi=JX_f$;
\item $\{f,h\}:=X_h(f)=\Omega(X_f,X_h)=\ev{-i[F,H]}$;
\item $g(X_f,X_h)=\ev{\tilde F\tilde H+\tilde H\tilde F}$; in particular $|X_f|^2=|\grad f|^2=2(\Delta F)^2$;
\item the geodesic distance is $d([\psi],[\phi])=\sqrt2\arccos|\ip{\psi}{\phi}|$, so the transition probability is $|\ip{\psi}{\phi}|^2=\cos^2\!\bigl(d/\sqrt2\bigr)$;
\item $[X_f,X_h]=-X_{\{f,h\}}$.
\end{enumerate}
\end{lemma}
\begin{proof}
For horizontal $v$, $df(v)=2{\rm Re}\ip{v}{\tilde F\psi}=g(v,\tilde F\psi)$, giving $\grad f=\tilde F\psi$, which is horizontal. With $X_f=-i\tilde F\psi$, $\Omega(X_f,v)=2{\rm Im}\,(i\ip{\tilde F\psi}{v})=df(v)$, proving (a). Then $\{f,h\}=g(\grad f,X_h)=2{\rm Re}\ip{\tilde F\psi}{-i\tilde H\psi}=-2{\rm Im}\ev{\tilde H\tilde F}=\ev{-i[F,H]}$, which is (b). Part (c) is $2{\rm Re}\ip{\tilde F\psi}{\tilde H\psi}$. For (d), the curve $s\mapsto[\cos s\,\psi+\sin s\,w]$, with $w\perp\psi$ a unit vector, is a geodesic of speed $\sqrt2$ that reaches $[\phi]$ at $s=\arccos|\ip{\psi}{\phi}|$ for suitable $w$. For (e), $X_F$ is the projection of the linear vector field $\psi\mapsto-iF\psi$, and the Lie bracket of linear fields $V_A,V_B$ is $V_{[B,A]}$; hence $[X_f,X_h]$ is the projection of $\psi\mapsto[F,H]\psi=-i(i[F,H])\psi$, i.e.\ $X_{\ev{i[F,H]}}=-X_{\{f,h\}}$.
\end{proof}

Part (b) is the geometric form of Ehrenfest's theorem: with $h=\ev H$, $\dot f=\{f,h\}=\ev{-i[F,H]}$ is Schr\"odinger evolution. Part (c) says that the length of $X_f$ measures the uncertainty of $F$. Part (d) corrects \cite{Corichi2008}, where the probability was written as $\cos^2d$; that expression corresponds to the Fubini--Study normalization $\tfrac12g$, not to the normalization in which (c) holds.

The Riemannian structure also gives gradient flows, which have no classical counterpart in the symplectic picture.

\begin{lemma}[Gradient flows are imaginary-time evolution]\label{lem:gradflow}
The negative gradient flow of $f=\ev F$, $\dot\psi=-\grad f$, through $[\psi_0]$ is $[\psi_t]=[e^{-tF}\psi_0]$.
\end{lemma}
\begin{proof}
For $\psi_t=e^{-tF}\psi_0/\|e^{-tF}\psi_0\|$ we get $\dot\psi_t=-F\psi_t+f(\psi_t)\psi_t=-\tilde F\psi_t$, using $\frac{d}{dt}\log\|e^{-tF}\psi_0\|=-f(\psi_t)$.
\end{proof}

\begin{lemma}[Morse--Bott property \cite{Frankel,Kirwan}]\label{lem:morsebott}
The critical set of $f=\ev F$ is the union of the submanifolds $\Ps(E_\lambda)$ for the eigenspaces $E_\lambda$ of $F$. At $[\psi]\in\Ps(E_\lambda)$ and for unit $v\perp\psi$,
\[
\Hess f(v,v)=2\,\bigl(\ip{v}{Fv}-\lambda\bigr),
\]
so the eigenvalues of $g^{-1}\Hess f$ are $0$ on $T\Ps(E_\lambda)$ and $\lambda'-\lambda$, with real multiplicity $2\dim E_{\lambda'}$, for $\lambda'\neq\lambda$. In particular $f$ is Morse--Bott, and the index of $\Ps(E_\lambda)$ is $2\sum_{\lambda'<\lambda}\dim E_{\lambda'}$.
\end{lemma}
\begin{proof}
Critical points are zeros of $\tilde F\psi$, i.e.\ eigenvectors. Along the geodesic $[\cos s\,\psi+\sin s\,v]$, $f=\lambda\cos^2s+\ip{v}{Fv}\sin^2s$, since the cross term $\sin2s\,{\rm Re}\ip{\psi}{Fv}=\lambda\sin2s\,{\rm Re}\ip{\psi}{v}$ vanishes. Differentiating twice and dividing by $g(v,v)=2$ gives the result.
\end{proof}

\section{Quantum first-class constraints: the discrete case}\label{sec:first-class}

In this section the constraints $\hat C_1,\dots,\hat C_M$ are self-adjoint operators and we assume that zero is an isolated point of the spectrum of the master operator $\Mop$ defined below, so that the physical space $V:=\bigcap_i\ker\hat C_i$ is a closed non-zero subspace and $\Mop$ has a spectral gap $\Delta>0$ above zero. This holds, for instance, whenever $\Hil$ is finite-dimensional and $V\neq0$. We write $\Pz$ for the orthogonal projection onto $V$ and $k=\dim V$.

\subsection{The quantum constraint function}

Following \cite{Corichi2008}, the naive condition $\ev{\hat C}=0$ is too weak: it has many solutions with $\hat C\Psi\neq0$. Instead, positivity of $\hat C^2$ gives $\ev{\hat C^2}=\|\hat C\Psi\|^2$, which vanishes if and only if $\hat C\Psi=0$. For several constraints, and a constant positive-definite $K^{ij}$, define the \emph{master operator} and the \emph{quantum constraint function}
\begin{equation}
\Mop:=K^{ij}\hat C_i\hat C_j,\qquad \Cq:=\ev{\Mop}:\Ps\to[0,\infty). \label{eq:Cq}
\end{equation}
Since $K$ is symmetric, $\Mop$ is self-adjoint (defined as a form sum), and $\Cq=K^{ij}{\rm Re}\ip{\hat C_i\psi}{\hat C_j\psi}\ge0$ vanishes exactly on $\Pphy:=\Ps(V)$. So the function that implements the Dirac conditions on $\Ps$ is \emph{the expectation value of the master constraint operator}. By choosing a basis orthonormal for $K$ we may take $K^{ij}=\delta^{ij}$ when convenient.

\begin{remark}[Exponentiated form]\label{rem:exp}
One may also impose $e^{i\lambda\hat C}\Psi=\Psi$ through $\Cq_\lambda:=\ev{1-\cos\lambda\hat C}\ge0$. For a \emph{single} $\lambda$ this is weaker than (\ref{eq:dirac}): it only forces $\spec\hat C\subset(2\pi/\lambda)\Z$ on the support of $\Psi$. The correct statement is $\Cq_\lambda=0$ for all $\lambda$, or for two incommensurate values. Unlike $\Cq$, the functions $\Cq_\lambda$ are bounded, and in Section~\ref{sec:continuous} this makes them the right objects when zero lies in the continuous spectrum.
\end{remark}

\subsection{The decomposition theorem}

Let $c_i:=\ev{\hat C_i}$ and $X_i:=X_{c_i}$. When the constraints close, $[\hat C_i,\hat C_j]=if_{ij}{}^k\hat C_k$, they generate a unitary action of a Lie group $G$, which acts on $\Ps$ by K\"ahler isometries. By Lemma~\ref{lem:dict}(b), $\{c_i,c_j\}=f_{ij}{}^kc_k$, so $\mu:=(c_i):\Ps\to\Lie^*$ is an equivariant moment map for this action, and the $X_i$ are its fundamental vector fields. The following identity does not require closure.

\begin{theorem}[Decomposition]\label{thm:decomp}
For any self-adjoint $\hat C_i$ and positive $K^{ij}$,
\begin{equation}
\Cq=K^{ij}c_ic_j+\tfrac12K^{ij}g(X_i,X_j)=:|\mu|_K^2+\tfrac12\|X_\mu\|_K^2. \label{eq:decomp}
\end{equation}
For a single constraint, $\Cq=c^2+\tfrac12|X_c|^2=c^2+\tfrac12|\grad c|^2$.
\end{theorem}
\begin{proof}
By symmetry of $K$, $\Cq=\tfrac12K^{ij}\ev{\hat C_i\hat C_j+\hat C_j\hat C_i}$, and by Lemma~\ref{lem:dict}(c), $\tfrac12\ev{\hat C_i\hat C_j+\hat C_j\hat C_i}=c_ic_j+\tfrac12g(X_i,X_j)$.
\end{proof}

\begin{corollary}\label{cor:fixed}
$\Pphy=\mu^{-1}(0)\cap\{X_1=\dots=X_M=0\}=\Ps\bigl(\bigcap_i\ker\hat C_i\bigr)$. If the $\hat C_i$ generate an action of a connected group $G$, then $\Pphy=\mu^{-1}(0)\cap\Ps^G$, where $\Ps^G$ is the fixed-point set of $G$ in $\Ps$.
\end{corollary}
\begin{proof}
Both terms in (\ref{eq:decomp}) are non-negative. $X_i([\psi])=0$ means $\hat C_i\psi\propto\psi$, and together with $c_i=0$ this gives $\hat C_i\psi=0$. For connected $G$, a point is fixed if and only if all fundamental vector fields vanish there.
\end{proof}

This makes precise the statement that one quantum step accomplishes the two classical ones. The first term of (\ref{eq:decomp}) is the quantum counterpart of the classical constraints, $\mu=0$. Classically one then identifies the points along the orbits of the $X_i$. The second term does something different: rather than identifying the points of a gauge orbit, it retains only those points whose orbit is a single point. There are therefore no gauge orbits left to quotient. In Section~\ref{sec:semiclassical} we shall see that the second term is of order $\hbar$ on semiclassical states, which is why it has no counterpart on $\Gb$.

The single-constraint case is instructive. The function $c=\ev{\hat C}$ is an admissible description of the real hypersurface $c^{-1}(0)$ (away from critical points), which is coisotropic with characteristic directions $X_c$, exactly like a classical first-class constraint surface. Theorem~\ref{thm:decomp} shows $\Pphy=c^{-1}(0)\cap\Crit(c)$: the Dirac condition selects the critical points of the moment map on its zero level.

\subsection{Geometry of the physical state space}

\begin{proposition}\label{prop:Pphy}
$\Pphy=\Ps(V)\cong\CP^{k-1}$ is a closed complex submanifold of $\Ps$ and:
\begin{enumerate}[(i)]
\item it is totally geodesic, and the induced structure makes it a K\"ahler manifold isometric to $\Ps(V)$ with the same normalization;
\item the pull-back of $\Omega$ to $\Pphy$ is non-degenerate: there are no gauge directions;
\item geodesic distances, hence transition probabilities (Lemma~\ref{lem:dict}(d)), between physical states computed intrinsically agree with those computed in $\Ps$;
\item the constraint functions $c_i$ and $\Cq$ generate nothing on $\Pphy$: $X_{c_i}=X_{\Cq}=0$ there.
\end{enumerate}
\end{proposition}
\begin{proof}
$T_{[\psi]}\Pphy=V\cap\psi^\perp$ is $J$-invariant, so $\Pphy$ is a complex submanifold; then $\Omega(v,Jv)=g(Jv,Jv)=g(v,v)\neq0$ for $v\neq0$ gives (ii). The unitary $U=\Pz-(1-\Pz)$ induces an isometry of $\Ps$ whose fixed-point set is $\Ps(V)\sqcup\Ps(V^\perp)$; connected components of fixed-point sets of isometries are totally geodesic, giving (i). For (iii), the minimizing geodesic between two points of $\Ps(V)$ lies in the projective line they span, which is contained in $\Ps(V)$. For (iv), $X_{c_i}=-i(\hat C_i-c_i)\psi=0$ and $X_{\Cq}=-i(\Mop-\Cq)\psi=0$ for $\psi\in V$.
\end{proof}

Proposition~\ref{prop:Pphy}(iv) was the observation of \cite{Corichi2008}; parts (i) to (iii) strengthen it. Classically $\Gb$ is coisotropic and the reduction is forced by the kernel of $\bar\Omega$; quantum mechanically $\Pphy$ is symplectic and no kernel exists to quotient by. Part (iii) is a consistency check on the probabilistic interpretation: the physical theory can compute probabilities without reference to the kinematical one.

\begin{proposition}[$\Cq$ is a quantum master constraint]\label{prop:morsebott}
\begin{enumerate}[(i)]
\item $\Cq$ is an inadmissible description of $\Pphy$: $d\Cq$ and $X_{\Cq}$ vanish on $\Pphy$, exactly as for the classical $\Mcl$ and unlike $c$.
\item $\Cq$ is Morse--Bott. $\Pphy$ is its absolute minimum, and the eigenvalues of $g^{-1}\Hess\Cq$ normal to $\Pphy$ are the non-zero eigenvalues $m$ of $\Mop$, each with real multiplicity $2\dim\ker(\Mop-m)$. In particular the spectral gap $\Delta$ of $\Mop$ is the smallest curvature of the potential well around $\Pphy$.
\item The other critical manifolds are $\Ps(\ker(\Mop-m))$ for $m>0$ in the spectrum of $\Mop$; the Morse index of each is $2\sum_{0\le m'<m}\dim\ker(\Mop-m')$.
\end{enumerate}
\end{proposition}
\begin{proof}
Apply Lemma~\ref{lem:morsebott} to $F=\Mop$ and Proposition~\ref{prop:Pphy}(iv).
\end{proof}

So the precise quantum analogue of the classical constraint functions $C_i$ is not $\ev{\hat C_i}$ but $\Cq$, and $\Cq$ is the analogue of the classical \emph{master} constraint. This explains why the master equation appears naturally in the quantum theory (Section~\ref{sec:observables}). Proposition~\ref{prop:morsebott} also covers the variant in which $\Mop$ has no kernel and one imposes $\Mop-m_0=0$ with $m_0=\min\spec\Mop$ \cite{Thiemann}. Then $\Cq-m_0$ has the same vector fields and $\Pphy$ is replaced by the absolute minimum $\Ps(\ker(\Mop-m_0))$. This suggests characterising physical states \emph{variationally}, as the minimum set of $\Cq$; we return to this in Section~\ref{sec:anomaly}.

\subsection{Dirac quantization versus K\"ahler reduction}

Suppose the constraints generate a unitary action of a compact connected group $G$. Then $\Ps$ is a Hamiltonian $G$-space with moment map $\mu$ and one could, mimicking the classical procedure, define the K\"ahler (Marsden--Weinstein) quotient
\[
\Ps/\!\!/G:=\mu^{-1}(0)/G.
\]
This is a stratified K\"ahler space, with strata labelled by orbit type \cite{SjamaarLerman}. By Kempf--Ness and Kirwan \cite{KempfNess,Kirwan} it is homeomorphic to the projective GIT quotient of $\Ps$ by the complexified group $G_\C$.

\begin{proposition}\label{prop:kahlerred}
$\Pphy$ consists of the points of $\mu^{-1}(0)$ with stabilizer $G$. Its image in $\Ps/\!\!/G$ is therefore the stratum of orbit type $G$, the deepest stratum, and the map $\Pphy\to\Ps/\!\!/G$ is injective. When $K$ is ${\rm Ad}$-invariant, $\Mop$ is a Casimir operator, $\Cq$ is $G$-invariant and descends to $\Ps/\!\!/G$, where it equals $\tfrac12\|X_\mu\|^2$ and its zero set is the image of $\Pphy$.
\end{proposition}
\begin{proof}
The first statement is Corollary~\ref{cor:fixed}. Orbits of fixed points are single points, so distinct physical states have distinct images. If $K$ is ${\rm Ad}$-invariant, $\Mop$ commutes with $G$, so $\Cq$ is invariant; on $\mu^{-1}(0)$ the first term in (\ref{eq:decomp}) vanishes.
\end{proof}

Thus Dirac quantization is not K\"ahler reduction of the quantum phase space. The quotient $\Ps/\!\!/G$ is generically much larger than $\Pphy$ (its real dimension is $\dim\Ps-2\dim G+2\dim H$, with $H$ the generic stabilizer, independently of $k$), and $\Pphy$ sits inside it as the most singular stratum. The two examples in Section~\ref{sec:examples} make this concrete. The comparison places the familiar ``quantization commutes with reduction'' question \cite{GuilleminSternberg} inside the quantum state space itself: the projectivization $\Ps(\Hil^G)$ of the invariant subspace is not the reduction of $\Ps$, but the fixed stratum of that reduction, singled out by the vanishing of the orbit-speed term.

\subsection{The quantum projection: gradient flow, group averaging and nearest points}

In \cite{Corichi2008} the group-averaging map was described as a ``true projection'' $\eta$ within $\Ps$, distinct from the classical quotient, but it was not characterised geometrically. The Riemannian structure of $\Ps$ now does this.

\begin{theorem}\label{thm:flow}
Let $\Delta>0$ be the spectral gap of $\Mop$ above zero.
\begin{enumerate}[(i)]
\item \emph{Gradient flow.} The negative gradient flow of $\Cq$ through $[\psi_0]$ is $[\psi_t]=[e^{-t\Mop}\psi_0]$.
\item \emph{Convergence.} If $\Pz\psi_0\neq0$, then $[\psi_t]\to\eta([\psi_0]):=[\Pz\psi_0]$, with
\[
1-|\ip{\hat\psi_t}{\hat\eta}|^2\le\frac{\|(1-\Pz)\psi_0\|^2}{\|\Pz\psi_0\|^2}\,e^{-2\Delta t},
\]
where hats denote unit vectors.
\item \emph{Nearest points.} $\eta([\psi])$ is the unique point of $\Pphy$ nearest to $[\psi]$, at distance $d([\psi],\Pphy)=\sqrt2\arccos\|\Pz\hat\psi\|$. The probability that a kinematical state passes the test ``is it physical?'' is $\|\Pz\hat\psi\|^2=\cos^2\bigl(d([\psi],\Pphy)/\sqrt2\bigr)$.
\item \emph{Fibration.} The basin of attraction of $\Pphy$ is the open dense set $\Ps\setminus\Ps(V^\perp)$, and $\eta$ is a holomorphic retraction onto $\Pphy$. Its fibres $\eta^{-1}([\phi])=\{[\phi+w]:w\in V^\perp\}$ are complex affine spaces of dimension $\dim V^\perp$, meeting $\Pphy$ orthogonally; $\Ps\setminus\Ps(V^\perp)$ is the total space of $\mathcal O(1)\otimes V^\perp\to\Ps(V)$.
\item \emph{Complexified gauge invariance.} If the $\hat C_i$ generate a representation of $G$ which extends to $G_\C$ (for instance $G$ compact, or $G=\R^k$ with $\Hil$ finite-dimensional), each fibre of $\eta$ is $G_\C$-invariant and is therefore a union of complexified gauge orbits.
\item \emph{Group averaging.} Let $G$ be compact or $G=\R^k$, let $\hat C_i$ correspond to an orthonormal basis of $\Lie$ for a bi-invariant metric, and let $p_t$ be the heat kernel of the corresponding Laplacian on $G$. Then
\begin{equation}
e^{-t\Mop}=\int_Gp_t(g)\,U(g)\,dg. \label{eq:heatavg}
\end{equation}
For $G=\R^k$ this reads $e^{-t\Mop}=(4\pi t)^{-k/2}\int_{\R^k}e^{-|\lambda|^2/4t}e^{i\lambda\cdot\hat C}d^k\lambda$. Hence the gradient flow of $\Cq$ is normalized heat-kernel-smeared group averaging. For compact $G$, $p_t\to1/{\rm vol}(G)$ and the $t\to\infty$ limit is Haar group averaging, $\int_GU(g)\,dg/{\rm vol}(G)=\Pz$. For $G=\R^k$ the limit coincides with the mean-ergodic average $\lim_{T\to\infty}(2T)^{-k}\int_{[-T,T]^k}U(\lambda)\,d^k\lambda=\Pz$ \cite{ReedSimon}.
\end{enumerate}
\end{theorem}
\begin{proof}
(i) is Lemma~\ref{lem:gradflow}. For (ii), write $e^{-t\Mop}\psi_0=\Pz\psi_0+R_t$ with $\|R_t\|\le e^{-\Delta t}\|(1-\Pz)\psi_0\|$ and $R_t\perp V$; the fidelity with $\hat\eta$ is $\|\Pz\psi_0\|^2/(\|\Pz\psi_0\|^2+\|R_t\|^2)$. For (iii), by Cauchy--Schwarz $|\ip{\hat\psi}{\phi}|\le\|\Pz\hat\psi\|$ for unit $\phi\in V$, with equality only for $\phi\propto\Pz\psi$; then use Lemma~\ref{lem:dict}(d). (iv) is the standard description of linear projection from $\Ps(V^\perp)$ onto $\Ps(V)$; the tangent space of a fibre at $[\phi]$ is $V^\perp$, orthogonal to $T\Pphy$. For (v), $U_\C(g)$ acts trivially on $V$ and preserves $V^\perp$, so $\Pz U_\C(g)\psi=\Pz\psi$. For (vi), let $\tilde X_i$ be the left-invariant fields and $\Delta_G=\sum_i\tilde X_i^2$. Writing $U(f)=\int fU\,dg$, unimodularity gives $U(\tilde X_if)=-U(f)\,dU(\tilde X_i)=-iU(f)\hat C_i$, so $U(\Delta_Gf)=-U(f)\Mop$. Hence $\frac d{dt}U(p_t)=-U(p_t)\Mop$ with $U(p_0)=1$, which gives (\ref{eq:heatavg}). For $G=\R^k$ this is the Fourier transform of a Gaussian, applied through the joint spectral theorem.
\end{proof}

Theorem~\ref{thm:flow} identifies three a priori different constructions: group averaging along the gauge orbit, which is linear in $\Hil$; the gradient flow of the quantum master constraint, which is intrinsic to the K\"ahler geometry of $\Ps$; and the nearest-point projection onto $\Pphy$. It also answers the question left open in \cite{Corichi2008} of which quotient replaces the classical one. A gauge orbit $\{U(g)\psi\}$ lies at constant distance from $\Pphy$, and its points are not physical unless the orbit is trivial; it is contained in a single fibre of $\eta$. Each fibre, however, is much larger: a complex affine space invariant under $G_\C$. The quantum reduction is the retraction along these holomorphic fibres, not the quotient by real gauge orbits.

Two further features are worth stressing. First, the gradient flow of $\Cq$ is defined for \emph{any} set of constraints, including those with structure functions that do not generate a group; group averaging is not. This is exactly the situation the master constraint programme was designed for, and the gradient flow gives it a geometric meaning. Second, the convergence rate in (ii) is the spectral gap of $\Mop$, i.e.\ the curvature of the well in Proposition~\ref{prop:morsebott}.

\subsection{Physical observables}\label{sec:observables}

As in the classical case, we call a self-adjoint $\hat O$ a physical observable if the unitary flow it generates preserves the physical state space. In \cite{Corichi2008} this condition was expanded in powers of $\lambda$,
\[
\ip{\Psi}{e^{-i\lambda\hat O}\Mop e^{i\lambda\hat O}\Psi}=\sum_{n\ge0}\frac{(-i\lambda)^n}{n!}\ip{\Psi}{[\hat O,\Mop]_{(n)}\Psi}=0\qquad\forall\Psi\in V,
\]
with $[\hat O,\Mop]_{(0)}=\Mop$ and $[\hat O,\Mop]_{(n+1)}=[\hat O,[\hat O,\Mop]_{(n)}]$, and the double commutator was identified as the first non-trivial condition. The next result shows that this condition is already sufficient.

\begin{proposition}\label{prop:obs}
For bounded self-adjoint $\hat O$ the following are equivalent:
\begin{enumerate}[(a)]
\item $e^{i\lambda\hat O}$ maps $\Pphy$ to itself for all $\lambda\in\R$;
\item the HVF $X_o$ of $o=\ev{\hat O}$ is tangent to $\Pphy$;
\item $\ev{[\hat O,[\hat O,\Mop]]}=0$ on $\Pphy$;
\item $\hat OV\subset V$;
\item $[\hat O,\Pz]=0$.
\end{enumerate}
Moreover, on $\Pphy$,
\begin{equation}
\ev{[\hat O,[\hat O,\Mop]]}=-2K^{ij}{\rm Re}\ip{\hat C_i\hat O\psi}{\hat C_j\hat O\psi}, \label{eq:qmastereq}
\end{equation}
which is the exact analogue of the classical identity (\ref{eq:mastereq}).
\end{proposition}
\begin{proof}
Expanding the double commutator and using $\Mop\psi=0$ for $\psi\in V$ gives (\ref{eq:qmastereq}), whose right-hand side is $-2\ev{\hat O\Mop\hat O}$ and vanishes if and only if $\hat O\psi\in V$; so (c)$\Leftrightarrow$(d). Since $\hat O$ is self-adjoint, $\hat OV\subset V$ implies $\hat OV^\perp\subset V^\perp$, so (d)$\Leftrightarrow$(e). Clearly (e)$\Rightarrow$(a)$\Rightarrow$(b). Finally, $X_o=-i(\hat O-o)\psi$ is tangent to $\Ps(V)$ at $[\psi]$ if and only if $\hat O\psi\in V$, so (b)$\Rightarrow$(d).
\end{proof}

Both the strong condition $[\hat O,\hat C_i]=0$ and the weak one $[\hat O,\hat C_i]=A_i{}^j\hat C_j$ imply (d), but (d) is much weaker than either. In fact there is a sharper difference from the classical theory:

\begin{corollary}[Compression]\label{cor:compression}
For any self-adjoint $\hat F$ on $\Hil$, $f|_{\Pphy}=\ev{\Pz\hat F\Pz}|_{\Pphy}$, and the intrinsic HVF of $f|_{\Pphy}$, computed with the induced K\"ahler structure, is the orthogonal projection of $X_f$ onto $T\Pphy$. Hence every kinematical operator induces a physical observable $\Pz\hat F\Pz$, and the algebra of physical observables acting on $\Pphy$ is $B(V)$.
\end{corollary}
\begin{proof}
For a complex submanifold the symplectic and metric orthogonal complements coincide, so the intrinsic HVF of a restricted function is the orthogonal projection of the ambient one. At $\psi\in V$, projecting $-i(\hat F-f)\psi$ onto $V\cap\psi^\perp$ gives $-i(\Pz\hat F\Pz-f)\psi$.
\end{proof}

Classically, $f|_{\Gb}$ is in general not gauge invariant and only invariant functions descend to $\Gh$. Quantum mechanically, $\Pphy$ is symplectic rather than coisotropic, so every function on it has a Hamiltonian flow, and restriction is always allowed. When zero lies in the continuous spectrum $\Pz=0$ and compression is unavailable; its role is then played by the RAQ requirement that observables commute with the rigging map (Section~\ref{sec:continuous}).

\subsection{Constraint algebra}

Classically the first-class relations (\ref{eq:firstclass}) imply that the $X_i$ close on $\Gb$, and on $\Gh$ they vanish altogether: the precise structure functions are not intrinsic, since suitable Darboux coordinates make the vector fields commute \cite{HT}. Quantum mechanically the vector fields relevant to $\Pphy$ are those of the functions that define it. For $\Cq_i:=\ev{\hat C_i^2}$, Lemma~\ref{lem:dict}(e) gives
\begin{equation}
[X_{\Cq_i},X_{\Cq_j}]=-X_{\ev{-i[\hat C_i^2,\hat C_j^2]}}, \label{eq:offshell}
\end{equation}
so the relevant off-shell objects are the commutators of the \emph{squares} of the constraints. All of these vector fields vanish on $\Pphy$. By Corollary~\ref{cor:fixed}, the question of whether the Dirac conditions are consistent is the question of whether the joint zero set of the $\Cq_i$ is non-empty; Section~\ref{sec:anomaly} gives a quantitative obstruction in terms of the commutators $[\hat C_i,\hat C_j]$ themselves.

\section{Examples}\label{sec:examples}

\subsection{A single U(1) constraint}

Let $\Hil=\C^3$ carry the spin-1 representation and $\hat C=\hat J_z$. Write $\psi=(a,b,c)$ in the basis $\ket{1,1},\ket{1,0},\ket{1,-1}$, so $\Ps=\CP^2$. Then
\[
c=\ev{\hat J_z}=|a|^2-|c|^2,\qquad \Cq=\ev{\hat J_z^2}=|a|^2+|c|^2=1-|b|^2,
\]
and $\Pphy=\{[\ket{1,0}]\}$ is a single point.

\emph{K\"ahler quotient.} $\mu^{-1}(0)=\{|a|=|c|\}$ is a real hypersurface in $\CP^2$. Modulo the gauge action $(a,b,c)\mapsto(e^{i\theta}a,b,e^{-i\theta}c)$ and the overall phase, the invariants are $|b|^2\in[0,1]$ and $\arg(ac\bar b^2)$, so $\Ps/\!\!/U(1)\cong S^2$. Its poles are $[\ket{1,0}]$ (stabilizer $U(1)$, the physical state) and $[(\ket{1,1}+\ket{1,-1})/\sqrt2]$ (stabilizer $\Z_2$). On this sphere $\Cq=1-|b|^2$ is, up to a constant, the moment map of the residual circle action $b\mapsto e^{i\alpha}b$: a height function whose minimum is $\Pphy$ and whose maximum is the orbifold point. The state $(\ket{1,1}+\ket{1,-1})/\sqrt2$ satisfies $\mu=0$, so it would be ``physical'' in the classical-mimicking reduction, but $\Cq=1$ there, entirely from the orbit-speed term $\tfrac12|X_c|^2=(\Delta J_z)^2$.

\emph{Gradient flow and observables.} The flow is $(a,b,c)\mapsto(e^{-t}a,b,e^{-t}c)$ normalized, with basin $\CP^2\setminus\Ps(V^\perp)$, the complement of the line $b=0$; on $\Ps/\!\!/U(1)$ it descends to the flow down the meridians. The spectral gap is $\Delta=1$. The physical observables are the operators with $\ket{1,0}$ as an eigenvector, for instance $\hat J_x^2+\hat J_y^2$; the compression of $\hat J_x$ is zero.

\subsection{Two spins with an SU(2) Gauss law}

Let $\Hil=\C^2\otimes\C^2$ and $\hat C_i=\hat J_i=\hat S^{(1)}_i+\hat S^{(2)}_i$, so $\Mop=\hat{\mathbf J}^2$ is the Casimir and $\Ps=\CP^3$. The only physical state is the singlet, so $\Pphy$ is a point. Every point of $\mu^{-1}(0)$, the set of states with $\ev{\hat{\mathbf S}^{(1)}}=-\ev{\hat{\mathbf S}^{(2)}}$, is $SU(2)$-equivalent to a point of the slice
\[
\psi_{\theta,\varphi}=\cos\theta\ket{\uparrow\downarrow}+e^{i\varphi}\sin\theta\ket{\downarrow\uparrow}.
\]
This follows from the Schmidt decomposition, using the fact that the reduced states of the two spins have Bloch vectors of equal length. The residual symmetry is the $\Z_2$ generated by a $\pi$ rotation about the $x$-axis, so $\Ps/\!\!/SU(2)\cong S^2/\Z_2$ is two-dimensional. On the slice,
\[
\Cq=\ev{\hat{\mathbf J}^2}=1+\sin2\theta\cos\varphi,
\]
again a height function. Its minimum $\Cq=0$ is the singlet ($\theta=\pi/4$, $\varphi=\pi$) and its maximum $\Cq=2$ is the triplet $\ket{1,0}$ ($\varphi=0$). The latter satisfies $\mu=0$ but is not physical: $\Cq=\tfrac12\|X_\mu\|^2=(\Delta J_x)^2+(\Delta J_y)^2=2$. The gradient flow $e^{-t\hat{\mathbf J}^2}$ damps the triplet component at rate $\Delta=2$, and by (\ref{eq:heatavg}) it is the $SU(2)$ heat-kernel average of the gauge rotations.

\section{Anomalies and second-class constraints}\label{sec:anomaly}

The decomposition theorem gives a geometric form of the uncertainty principle that controls when the Dirac conditions can be solved.

\begin{theorem}\label{thm:anomaly}
For self-adjoint $\hat C_1,\hat C_2$ and $\psi$ in their common domain,
\begin{equation}
\Cq_1+\Cq_2\ \ge\ c_1^2+c_2^2+|\{c_1,c_2\}|\ \ge\ |\{c_1,c_2\}|=\bigl|\ev{-i[\hat C_1,\hat C_2]}\bigr|. \label{eq:anomalybound}
\end{equation}
\end{theorem}
\begin{proof}
By Lemma~\ref{lem:dict}, $|\{c_1,c_2\}|=|\Omega(X_1,X_2)|=|g(JX_1,X_2)|\le|X_1||X_2|\le\tfrac12(|X_1|^2+|X_2|^2)$. Now use (\ref{eq:decomp}).
\end{proof}

\begin{corollary}[Central obstruction and Gupta--Bleuler]\label{cor:GB}
Suppose $[\hat C_1,\hat C_2]=i\kappa$ with $\kappa>0$, in the sense that $(\hat C_1,\hat C_2)$ satisfy the Weyl relations. Then $\Cq_1+\Cq_2\ge\kappa$ on $\Ps$, so there are no Dirac states. The bound is attained: with $\hat a:=(\hat C_1+i\hat C_2)/\sqrt{2\kappa}$, which satisfies $[\hat a,\hat a^\dagger]=1$,
\[
\hat C_1^2+\hat C_2^2=\kappa\,(2\hat a^\dagger\hat a+1),\qquad \Cq_1+\Cq_2=\kappa+2\kappa\|\hat a\psi\|^2,
\]
so the minimum set of $\Cq_1+\Cq_2$ is exactly the Gupta--Bleuler space $\{\hat a\psi=0\}$ \cite{Gupta,Bleuler}. By the Stone--von Neumann theorem $\Hil\cong L^2(\R)\otimes\mathcal K$ with $\hat a$ acting on the first factor, and the minimum set is $\Ps(\ket0\otimes\mathcal K)\cong\Ps(\mathcal K)$: the expected reduced state space after eliminating the second-class pair.
\end{corollary}
\begin{proof}
A direct computation: $\hat a^\dagger\hat a=(2\kappa)^{-1}(\hat C_1^2+\hat C_2^2+i[\hat C_1,\hat C_2])=(2\kappa)^{-1}(\hat C_1^2+\hat C_2^2-\kappa)$.
\end{proof}

Several remarks are in order. First, a central term is a \emph{symplectic} obstruction: it is the constant value of $\Omega(X_1,X_2)$, which bounds $\Cq$ from below through the orbit-speed term. Second, the condition $\mu=0$ alone (here $\ev{\hat C_1}=\ev{\hat C_2}=0$) has plenty of solutions; it is the orbit-speed term that cannot vanish. Third, Corollary~\ref{cor:GB} suggests a single variational principle covering all cases considered so far:
\begin{quote}
\emph{the physical state space is the minimum set of the quantum constraint function $\Cq$, equivalently the $\omega$-limit set of its gradient flow for generic initial data.}
\end{quote}
For first-class constraints with zero in the discrete spectrum this is $\Pphy$; with the shifted master constraint it is $\Ps(\ker(\Mop-m_0))$; for a second-class pair it is the Gupta--Bleuler space. Changing the weights $K^{ij}$ for a second-class pair replaces the vacuum by a squeezed vacuum, so the embedding of the minimum set depends on $K$, but its intrinsic K\"ahler geometry $\Ps(\mathcal K)$ does not. For second-class constraints that are not of Weyl type the minimum need not exist or need not be the physically correct answer, and this principle should be tested further.

This also sharpens the speculation in \cite{Corichi2008} about gauge versus symmetry anomalies. For generators of a symmetry, a central extension is harmless: it only makes the representation projective, and $\Ps$ is the natural home of projective representations. For constraints, the same central term is fatal to the Dirac conditions, by (\ref{eq:anomalybound}). The distinction is between functions that are required to vanish and functions that are only required to generate motions. Whether the Gupta--Bleuler-type minimisers give a consistent quantization in genuinely anomalous field theories, as with the highest-weight conditions of the Virasoro algebra, is an interesting open question.

\section{Zero in the continuous spectrum}\label{sec:continuous}

The previous sections assumed that zero is an isolated eigenvalue, a condition that, as noted in \cite{Corichi2008}, fails in many systems of interest. We now show that the geometric picture extends.

\subsection{Dirac states as zeros of the quantum constraint function}

Let $\mathfrak A\subset B(\Hil)$ be a unital C$^*$-algebra containing the constraint unitaries $U_i(\lambda)=e^{i\lambda\hat C_i}$, and $S(\mathfrak A)$ its state space with the weak-$*$ topology. Vector states of $\Ps$ embed in $S(\mathfrak A)$, and normal states are weak-$*$ dense. The bounded functions $\Cq_{i,\lambda}(\omega):=\omega(1-\cos\lambda\hat C_i)$ of Remark~\ref{rem:exp} extend continuously to all of $S(\mathfrak A)$.

\begin{definition}
The \emph{physical state space} is $\Sphy:=\{\omega\in S(\mathfrak A):\Cq_{i,\lambda}(\omega)=0\ \forall i,\lambda\}$.
\end{definition}

Since $\omega(1-\cos\lambda\hat C)=0$ forces ${\rm Re}\,\omega(U(\lambda))=1$ and $|\omega(U(\lambda))|\le1$, we have $\Sphy=\{\omega:\omega(U_i(\lambda))=1\ \forall i,\lambda\}$. These are precisely the \emph{Dirac states} of Grundling and Hurst \cite{GrundlingHurst}. $\Sphy$ is a weak-$*$ closed face of $S(\mathfrak A)$, and its normal part consists of the density matrices supported on $V$, so it reduces to (the convex hull of) $\Pphy$ in the discrete case. When $V=0$, $\Sphy$ consists of non-normal states, a zero set ``at infinity'' of $\Ps$.

\begin{theorem}[The gradient flow reaches Dirac states]\label{thm:dirac-states}
Let $\Mop=\sum_i\hat C_i^2$ (form sum), $\psi_0$ a unit vector, and suppose $0\in\supp\mu_{\psi_0}$, the spectral measure of $\Mop$ in $\psi_0$. Let $\omega_t$ be the vector states of $\psi_t=e^{-t\Mop}\psi_0/\|e^{-t\Mop}\psi_0\|$. Then:
\begin{enumerate}[(i)]
\item $\Cq(\psi_t)$ is non-increasing and tends to $0$;
\item every weak-$*$ cluster point $\omega$ of $\{\omega_t\}_{t\to\infty}$, and such points exist, is a Dirac state;
\item if $\Pz\psi_0\neq0$, the limit is the normal state $\eta([\psi_0])$ of Theorem~\ref{thm:flow}; if $\Pz\psi_0=0$, $\psi_t\rightharpoonup0$ weakly and every cluster point annihilates the compact operators.
\end{enumerate}
\end{theorem}
\begin{proof}
(i) Monotonicity holds because $\psi_t$ is a gradient flow. Write $\Cq(\psi_t)=\int xe^{-2tx}d\mu/\int e^{-2tx}d\mu$. Fix $\eps>0$. For $t\ge1/(2\eps)$ the function $xe^{-2tx}$ is decreasing on $[\eps,\infty)$, so
\[
\Cq(\psi_t)\le\eps+\frac{\eps\,e^{-2t\eps}}{e^{-t\eps}\mu([0,\eps/2])}\xrightarrow[t\to\infty]{}\eps,
\]
using $\mu([0,\eps/2])>0$. (ii) Cluster points exist by Banach--Alaoglu. From $1-\cos\lambda x\le\lambda^2x^2/2$ and $\hat C_i^2\le\Mop$ as forms, $\omega_t(1-\cos\lambda\hat C_i)\le\tfrac{\lambda^2}{2}\Cq(\psi_t)\to0$, and evaluation is weak-$*$ continuous. (iii) The first case is Theorem~\ref{thm:flow}. If $\Pz\psi_0=0$, let $E_\eps$ be the spectral projection of $\Mop$ on $[0,\eps]$. The argument in (i) gives $\|(1-E_\eps)\psi_t\|\to0$, so for any $\phi$, $\limsup|\ip{\phi}{\psi_t}|\le\|E_\eps(1-\Pz)\phi\|+|\ip{\Pz\phi}{\psi_t}|$. The last term vanishes because $\Pz e^{-t\Mop}\psi_0=\Pz\psi_0=0$, and the first tends to $0$ as $\eps\to0$. Weak convergence to zero implies $\omega_t(K)\to0$ for compact $K$.
\end{proof}

Theorem~\ref{thm:dirac-states} turns the geometric construction of Section~\ref{sec:first-class} into a construction of physical states that does not require zero to be an eigenvalue. The flow of the quantum master constraint, defined intrinsically on $\Ps$, carries every kinematical state whose spectral support reaches zero to the zero set of the constraint functions on the closure of $\Ps$.

\subsection{The rigging map from the gradient flow}

Dirac states alone do not provide an inner product. RAQ supplies one through the rigging map. Remarkably, it is encoded in the \emph{rate} at which the gradient flow escapes $\Ps$.

\begin{theorem}\label{thm:rigging}
Let $\hat C_1,\dots,\hat C_k$ be strongly commuting, $U(\lambda)=e^{i\lambda\cdot\hat C}$ and $\Mop=\sum_i\hat C_i^2$. Let $D\subset\Hil$ be a dense subspace such that $\lambda\mapsto\ip{\phi}{U(\lambda)\psi}$ is integrable on $\R^k$ for $\phi,\psi\in D$ (the standard RAQ hypothesis \cite{ALMMT,Marolf}). Then for $\phi,\psi\in D$,
\begin{equation}
\eta_{\rm RAQ}(\psi)[\phi]:=\int_{\R^k}\ip{\phi}{U(\lambda)\psi}\,d^k\lambda=\lim_{t\to\infty}(4\pi t)^{k/2}\,\ip{\phi}{e^{-t\Mop}\psi}. \label{eq:rigging}
\end{equation}
Consequently the physical inner product $\ip{\eta\phi}{\eta\psi}_{\rm phys}=\eta_{\rm RAQ}(\psi)[\phi]$ is the leading coefficient of the decay of kinematical overlaps along the gradient flow of $\Cq$:
$\ip{\phi}{e^{-t\Mop}\psi}\sim(4\pi t)^{-k/2}\ip{\eta\phi}{\eta\psi}_{\rm phys}$.
\end{theorem}
\begin{proof}
By (\ref{eq:heatavg}), $(4\pi t)^{k/2}\ip{\phi}{e^{-t\Mop}\psi}=\int e^{-|\lambda|^2/4t}\ip{\phi}{U(\lambda)\psi}\,d^k\lambda$. The integrand is dominated by $|\ip{\phi}{U(\lambda)\psi}|\in L^1$ and converges pointwise, so dominated convergence applies.
\end{proof}

The power of $t$ detects the spectral type at zero. If the joint spectral measure $\mu_{\phi\psi}$ of $\hat C$ consists of a point mass at the origin plus an absolutely continuous part with density $\rho_{\phi\psi}$ continuous at $0$, then
\[
\ip{\phi}{e^{-t\Mop}\psi}=\ip{\phi}{\Pz\psi}+\Bigl(\frac{\pi}{t}\Bigr)^{k/2}\rho_{\phi\psi}(0)+o(t^{-k/2}),
\]
and $(2\pi)^k\rho_{\phi\psi}(0)=\eta_{\rm RAQ}(\psi)[\phi]$. The discrete and continuous constructions are thus two orders of one asymptotic expansion. For compact $G$ the analogous statement is Theorem~\ref{thm:flow}(vi). For non-compact non-abelian groups, (\ref{eq:heatavg}) still holds for unimodular $G$, but the large-$t$ behaviour of $p_t$ is governed by the bottom of the spectrum of the Laplacian on $G$ and need not reduce to plain group averaging. This is precisely where group averaging is known to be delicate \cite{LoukoRovelli}, and the heat-kernel prescription may offer a natural regularization. We leave this for future work.

\subsection{Example: $\hat C=\hat p$}

On $\Hil=L^2(\R)$ let $\hat C=\hat p$, so $V=0$. With the unitary Fourier transform $\tilde\psi$,
\[
\ip{\phi}{e^{-t\hat p^2}\psi}=\int\overline{\tilde\phi(k)}\tilde\psi(k)e^{-tk^2}dk\sim\sqrt{\pi/t}\;\overline{\tilde\phi(0)}\tilde\psi(0),
\]
so (\ref{eq:rigging}) gives $\eta_{\rm RAQ}(\psi)[\phi]=2\pi\overline{\tilde\phi(0)}\tilde\psi(0)=\overline{\textstyle\int\phi}\int\psi$, the standard result: the physical Hilbert space is one-dimensional, spanned by the constant function. Along the normalized flow, $|\tilde\psi_t(k)|^2\propto e^{-2tk^2}|\tilde\psi(k)|^2$ concentrates at $k=0$, the wave packet spreads, $\Cq(\psi_t)\to0$, $\psi_t\rightharpoonup0$, and $\omega_t(e^{i\lambda\hat p})\to1$. The cluster points are translation-invariant Dirac states. The same analysis applies directly to the parametrized particle and to loop quantum cosmology with a massless scalar field \cite{APS}, where it would be interesting to compare the heat-flow construction with the group averaging used there.

\section{Semiclassical limit and the classical master constraint}\label{sec:semiclassical}

We restore $\hbar$ and pull $\Cq$ back to classical phase space along coherent states. Let $\Gamma=\R^{2N}$ with canonical coordinates $z=(q,p)$, let $\ket z$ be the standard (isotropic) coherent states, and let $\hat C$ be the Weyl quantization of a polynomial classical constraint $C$.

\begin{proposition}\label{prop:semiclassical}
With $|\nabla C|^2$ and $\Delta$ the Euclidean gradient norm and Laplacian on $\R^{2N}$,
\begin{equation}
\Cq(z):=\ip{z}{\hat C^2z}=C^2+\frac\hbar2\bigl(|\nabla C|^2+C\,\Delta C\bigr)+O(\hbar^2). \label{eq:semiclassical}
\end{equation}
Term by term, the decomposition (\ref{eq:decomp}) becomes $c(z)^2=C^2+\tfrac\hbar2C\Delta C+O(\hbar^2)$ and $(\Delta\hat C)_z^2=\tfrac\hbar2|\nabla C|^2+O(\hbar^2)$. On $\Gb$,
\[
\Cq\big|_{\Gb}=\frac\hbar2|X_C|^2+O(\hbar^2),
\]
where $|X_C|$ is the norm of the classical gauge vector field in the K\"ahler metric of the coherent-state complex structure.
\end{proposition}
\begin{proof}
The coherent-state expectation value (Husimi symbol) of an operator is $e^{(\hbar/4)\Delta}$ applied to its Weyl symbol \cite{Folland}. The Weyl symbol of $\hat C^2$ is $C\star C=C^2+O(\hbar^2)$, because the $O(\hbar)$ term $\tfrac{i\hbar}{2}\{C,C\}$ vanishes. Hence $\Cq=C^2+\tfrac\hbar4\Delta(C^2)+O(\hbar^2)$, and $\Delta(C^2)=2|\nabla C|^2+2C\Delta C$. Similarly $c=C+\tfrac\hbar4\Delta C+O(\hbar^2)$. Finally $|\nabla C|=|X_C|$ for the flat K\"ahler structure.
\end{proof}

This proposition makes the relation with the master constraint programme a theorem. The classical master constraint $\Mcl=C^2$ is the $\hbar\to0$ limit of $\Cq$, so the quantum double-commutator condition of Proposition~\ref{prop:obs} has the master equation (\ref{eq:mastereq}) as its classical limit. In \cite{Corichi2008} we stressed that we had \emph{not} quantized (\ref{eq:mastereq}); we now see that it is the classical limit of the Dirac condition. Moreover the leading quantum correction is precisely the orbit-speed term: since $C$ is an admissible description, $|X_C|\neq0$ on $\Gb$, so no coherent state is physical. Semiclassical states sit at height $\tfrac\hbar2|X_C|^2$ in the potential well of Proposition~\ref{prop:morsebott}, and physical semiclassical states are obtained by applying $\eta$ to them, in agreement with \cite{ABC}.

\begin{example}
For linear $C=\alpha q+\beta p$, (\ref{eq:semiclassical}) is exact: $\Cq=C^2+\tfrac\hbar2(\alpha^2+\beta^2)$. For $C=\tfrac12(q^2+p^2)-E$, $\hat C=\hbar(\hat N+\tfrac12)-E$, so $V\neq0$ if and only if $E=\hbar(n+\tfrac12)$, the Bohr--Sommerfeld condition with its Maslov correction. On coherent states over the classical circle $\Gb$, one finds exactly $\Cq=\hbar E+\hbar^2/4$, with $\hbar E=\tfrac\hbar2|\nabla C|^2$. The probability that such a state is physical is $\|\Pz\ket z\|^2=e^{-\bar n}\bar n^{\,n}/n!$, where $\bar n=n+\tfrac12$; for large $n$ it behaves as $(2\pi n)^{-1/2}$. The reduced phase space is a point, and the physical state is the single eigenstate $\ket n$.
\end{example}

\section{Discussion}\label{sec:discussion}

We have revisited the Dirac quantization of constrained systems from the perspective of the K\"ahler geometry of the quantum state space. In \cite{Corichi2008} we showed that the Dirac conditions define the zero set of a function on $\Ps$ and that no degenerate gauge directions appear. Here we have explained why. The quantum constraint function is the expectation value of the master operator, and it decomposes into a moment-map term and an orbit-speed term (Theorem~\ref{thm:decomp}). The first reproduces the classical constraints; the second collapses the gauge orbits rather than quotienting them. The physical state space is the fixed stratum of the K\"ahler quotient, and the quantum ``quotient'' is a holomorphic retraction whose fibres are invariant under the complexified gauge group, realised by the gradient flow of $\Cq$, which coincides with heat-kernel group averaging. The double-commutator condition on observables, which we had found as the first term of an expansion, is exact.

The geometric viewpoint pays off most clearly beyond the simplest case. The uncertainty bound (\ref{eq:anomalybound}) turns central terms into symplectic obstructions and yields Gupta--Bleuler states as minimisers. For zero in the continuous spectrum the same gradient flow produces Dirac states and, through its decay rate, the RAQ physical inner product. The semiclassical expansion shows that the classical master constraint is the leading term of $\Cq$ and identifies the first quantum correction with the gauge-orbit speed.

Several directions remain open.
\begin{enumerate}[(a)]
\item \emph{Structure functions.} The gradient flow of $\Cq$ requires no group. For constraint algebras with structure functions, such as the hypersurface-deformation algebra of general relativity, it offers a candidate for the projection onto physical states where group averaging is unavailable. Its convergence properties, controlled by the spectrum of $\Mop$ near zero, deserve study.
\item \emph{Non-compact non-abelian groups.} The large-time asymptotics of (\ref{eq:heatavg}) for semisimple groups should be compared with the known subtleties of group averaging \cite{LoukoRovelli}.
\item \emph{Applications.} Loop quantum cosmology with a massless scalar \cite{APS} and the parametrized particle are natural testing grounds for Theorem~\ref{thm:rigging}, including a comparison of physical semiclassical states obtained by the flow with those of \cite{ABC}.
\item \emph{BRST.} The Morse--Bott structure of $\Cq$, with $\Pphy$ as minimum and the spectrum of $\Mop$ as transverse Hessian, resembles the role of a BRST Laplacian whose kernel represents cohomology, and Witten's deformation of Morse theory \cite{Witten}. Making this precise would connect the present approach with \cite{BarnichGrigoriev}.
\item \emph{Variational principle.} The proposal of Section~\ref{sec:anomaly}, that physical states minimize $\Cq$, should be tested on non-Weyl second-class systems and on anomalous field theories.
\end{enumerate}

We hope that this geometric viewpoint, besides reproducing the standard conditions, sheds light on the conceptual challenges of constrained quantization and, in particular, of quantum gravity.

\section*{Acknowledgments}
I would like to thank A.~Ashtekar, K.~Giesel and J.~A.~Zapata for discussions on the original version. That work was in part supported by CONACyT U47857-F grant, by NSF PHY04-56913, and by the Eberly Research Funds of Penn State.

\appendix
\section{Changes with respect to the published version}\label{app:changes}

For the reader comparing with \cite{Corichi2008}:
\begin{enumerate}[(1)]
\item The $n=0$ term of the finite-flow expansion is $\{g,f\}_{(0)}=g$ (it read $f$); see (\ref{eq:finiteflow}).
\item With the normalization in which $(f,f)=2(\Delta F)^2$, the transition probability is $\cos^2(d/\sqrt2)$, not $\cos^2d$ (Lemma~\ref{lem:dict}(d)).
\item The condition $e^{i\hat C}\Psi=\Psi$, equivalently $\ev{1-\cos\hat C}=0$, for a single parameter value is not equivalent to $\hat C\Psi=0$; it must be imposed for all $\lambda$ (Remark~\ref{rem:exp}).
\item The double-commutator condition is not only the first non-trivial condition: on physical states it implies all higher ones (Proposition~\ref{prop:obs}).
\item Index typos in the commutator relations have been corrected and the sign fixed: $[X_{\Cq_i},X_{\Cq_j}]=-X_{\ev{-i[\hat C_i^2,\hat C_j^2]}}$ (\ref{eq:offshell}).
\item The statement that the Dirac conjecture is ``completely justified'' has been qualified (footnote in Section~\ref{sec:classical}).
\end{enumerate}

\end{document}